# *sp*-Electron Magnetic Clusters with a Large Spin in Graphene

Danil W. Boukhvalov[1,2] and Mikhail I. Katsnelson[3]

[1] *School of Computational Sciences, Korea Institute for Advanced Study (KIAS) Hoegiro 87, Dongdaemun-Gu, Seoul, 130-722, Korean Republic*

[2] *Computational Materials Science Center, National Institute for Materials Science, 1-2-1 Sengen, Tsukuba, Ibaraki 305-0047, Japan*

[3] *Radboud University Nijmegen, Institute for Molecules and Materials, Heyendaalseweg 135, 6525 AJ Nijmegen, the Netherlands*

**ABSTRACT** *Motivated by recent experimental data (*Sepioni, M. *et al. Phys. Rev. Lett.* **2010**, *105*, 207205*), we have studied the possibility of forming magnetic clusters with spin $S>½$ on graphene by adsorption of hydrogen atoms or hydroxyl groups. Migration of hydrogen atoms and hydroxyl groups on the surface of graphene during the delamination of HOPG led to the formation of seven-atom or seven-OH-group clusters with $S=5/2$ that were of a special interest. The coincidence of symmetry of the clusters with the graphene lattice strengthens the stability of the cluster. For $(OH)_7$ clusters that were situated greater than 3 nm from one another, the reconstruction barrier to a nonmagnetic configuration was approximately 0.4 eV, whereas for $H_7$ clusters, there was no barrier and the high-spin state was unstable. Stability of the high-spin clusters increased if they were formed on top of ripples. Exchange interactions between the clusters were studied and we have shown that the ferromagnetic state is improbable. The role of the chemical composition of the solvent used for the delamination of graphite is discussed.*

E-mail: [danil@kias.re.kr](danil@kias.re.kr)

**KEYWORDS:** graphene, sp-magnetism, graphene functionalization, graphene oxide

Graphene is considered to be a prospective material for future electronics.[1-4] Its magnetic properties have attracted attention due to their potential application in spintronics[5] and due to the importance of the problem of *sp*-electron magnetism, which is interesting both conceptually and as a route to generating new magnetic semiconductors with high Curie temperatures.[6] *sp*-adsorbates such as hydrogen,[10,11] vacancies,[12,13] and 3*d* and 4*f* elements adsorbed on graphene[14-15] were considered as the source of the magnetic moment in graphene edges.[7-9] It is unclear how robust the magnetism is and if it can be used in practice. For the magnetism of the edges, there are two major problems, namely, oxidation[16-21] and edge reconstruction,[22-24] which both result in the disappearance of the magnetic moment. As for magnetism that is related to hydrogen impurities, vacancies and broken bonds of the bulk in general, the magnetic moment can be killed by a passivation of the broken bonds or a reconstruction of magnetic configurations into nonmagnetic ones.[18, 26, 27] Transition metals can destroy graphene[28] because their atoms demonstrate a strong tendency for clusterization,[29-33] which changes the magnetism in an uncontrollable way. Thus, it is not easy to create robust and controllable magnetism in graphene-based systems, and this is one of the greatest challenges in the study of the physical chemistry of carbon materials.

There have been several experimental studies that have observed paramagnetism in graphene[34] and magnetic ordering in related systems.[35-37] The efforts by Enoki *et al*[34]. is especially interesting for the following reasons: (1) pure graphene (more precisely, graphene laminate) was studied instead of reduced graphite oxide, nanographites or graphite; (2) the method used to produce the samples in this study[38,39] is one of the most promising methods for broad-scale industrial production of graphene, and the graphene laminate deserves special attention; and (3) the result of the experiment[34] produced a stable and reproducible magnetic moment of 4.5±0.5 $\mu_B$, which was unexpected in light of existing theoretical views on magnetism of carbon materials (for review, see recent works[11,18] and references therein).

The high-spin state was unexpected because experimental evidence[40] and theory[17,32,26] have previously shown that all possible *3d* impurities in graphene (adsorbates and substitution impurities) were in a low-spin state that was caused by the specific carbon environment (a flat geometry instead of octahedral or pyramidal geometry), and the magnetic moment was smaller than 3 $\mu_B$. However, if the observed paramagnetism[34] is related to pollution by 3d elements, a broad distribution of the magnetic moment from 1 to 3 $\mu_B$ would have been expected. It is worth noting that migration barriers for the *3d* atoms on graphene are typically small,[14-17,32,33] and are not stable, and iron impurities strengthen interlayer coupling in graphene, which suppresses the exfoliation of single layers.[32] The pollution by *4f* elements is very improbable and their concentration should be extremely low. Thus, the observed magnetism is likely related to *sp*-electrons. At the same time, existing magnetism models (vacancies) cannot explain the stability of magnetic centers with 4 or 5 $\mu_B$ that have been experimentally observed.[34] Adsorption of hydrogen or other covalently bound impurities to the carbon atoms of one of the sublattices[10] can result in formation of magnetic chains or clusters with a magnetic moment of 3 $\mu_B$ or greater of ferromagnetic[41] or antiferromagnetic[42] ordering with a magnetic moment of approximately 1 $\mu_B$ per two carbon atoms. The magnetic moment at vacancies is restricted to 2 $\mu_B$.[12,13] For zigzag edges, one could expect a broad distribution in the sizes of magnetic clusters and magnetic moments; however, this is in clear contradiction to observed experimental data[34], which shows a well-defined value for the magnetic moment.

In this work, we propose a model that explains this result. The model is based on a demonstration of the metastability of certain highly symmetric clusters of $(OH)_7$-groups with the desired magnetic properties. We studied the effects of different solvents on the formation of the cluster.

**A MODEL OF MAGNETIC CLUSTERS**

Yazyev and Helm[10] considered an interesting situation where one hydrogen atom was chemisorbed at one of the carbon sublattices and to two other atoms in another sublattice (Fig. 2a). They found that the total magnetic moment of this configuration was approximately 1 $\mu_B$ (two spins ½ up and one spin ½ down), and the unpaired electron was distributed throughout the whole triangle. If one adds one more hydrogen atom to the second sublattice, the total magnetic moment of the cluster increased to 2 $\mu_B$. Typically, such clusters are unstable.[43] However, the situation can be different for adatoms with a stronger bond to graphene, such as fluorine[18] or the OH group.[44] In addition, the geometry and symmetry of the graphene lattice is relevant.

Graphene laminate is produced from a solution, and the solvent is essential because the observed magnetic behavior[34] is sensitive to the type of the solvent used.[45] Based on the chemical structure of the dimethylformamide (DMF, $(CH_3)_2NC(O)H$, see Fig. 3a) and methylpyrrolidone (NMP, $C_5H_6NO$, see Fig. 3b) solvents, one can assume that a transfer of hydrogen and oxygen atoms to the surface of graphene is possible upon sonication. The oxygen atoms form nonmagnetic epoxy groups,[42] whereas hydrogen atoms and hydroxyl groups can be a potential source for the generation of a magnetic moment. To estimate how probable these transfer processes are, we have calculated the formation energy for single H and OH groups using the formula $E_{form} = E_{graphene+X} - (E_{graphene} + E_{solvent})$, where $E_{graphene}$ is the total energy of pure graphene, $E_{solvent}$ is the energy of a molecule of solvent in empty space, $E_{graphene+X}$ is the energy of graphene with a chemisorbed hydrogen atom or hydroxyl group plus the energy of one molecule of solvent without one hydrogen atom, or without one hydrogen and one oxygen atom (see Fig. 3c). For DMF, the formation energy for the hydrogen atom and the OH group was 5.24 eV and 5.73 eV, respectively. For NMP, the energies were 2.88 and 7.09 eV, respectively. To understand how big or small these energies are, one can compare them with the vacancy formation energy, which is 7 eV [13] and can be used as an upper limit for the strength of carbon-carbon bonds. For the case of multivacancies, the energy per removed carbon atom is smaller, approximately 4 eV,[46] which can be used as a lower limit. The destruction of these bonds takes place on sonication. The size of graphene flakes in graphene laminate are smaller than in HOPG, which means that rather large energies are involved in the exfoliation process used to produce graphene laminate.[38,39] One can assume that when using DMF as a solvent, the appearance of both H and OH groups at the surface of graphene is not excluded, whereas for the case of NMP, only hydrogen groups can be involved. It is worth noting that the atoms of oxygen and hydrogen migrate to graphene from DMF molecules that are situated nearby, which allows for the formation of OH groups. The large energy produced during the sonication process allows the samples to reach configurations of the hydroxyl groups that cannot be created in graphene oxide that is produced at room temperature[43,47] due to high-energy barriers. To study the role of solvent more systematically, we have considered chloroform as another solvent.[38,39] The migration energies for hydrogen and chlorine turn out to be 3.34 eV and 5.96 eV, respectively. Therefore, only the first case should be considered. According to our calculations,[48] both atomic and molecular chlorine is weakly bound with graphene and does not result in the appearance of a magnetic moment.

In principle, there are several combinations of chemisorbed species with spins that are larger than 1/2: (1) one H or OH is chemisorbed by a carbon atom from sublattice A and three other H or OH groups are chemisorbed by three nearest neighbors of the atom that belong to sublattice B; (2) or *six* other H or OH sit at second, third or fourth neighbors belonging to sublattice B (Fig. 2b-d). The second option is interesting because the total spin of the cluster is expected to be 5/2 (6/2-1/2) in all cases, which is in agreement with the experimental data.[34] To study this case quantitatively, we have performed the total-energy calculations for different configurations, as shown in Fig. 2b-d for cases when all hydrogen atoms or hydroxyl groups are on the same side of the graphene plane and for the case when the central group is on one side and all other groups are on the opposite side.

The configuration with all hydrogen adatoms or OH-groups on the same side is always more energetically favorable and has an energy of gain approximately 0.5~0.8 eV per group. The cluster shown in Fig. 2c has the lowest energy in comparison with the configuration of Fig. 2b (the energy gain is 2.2 and 2.8 eV for H and OH groups, respectively) and Fig. 2d (the energy gain is 1.4 and 1.8 eV for H and OH groups, respectively). The difference in energy is due to a strong dependence on the chemisorption energy for the graphene plane distortions,[50] which is especially strong for the hydroxyl groups.[44] Thus, the formation of the clusters that are shown in Fig. 2c are the most probable; therefore, we will only consider this configuration. It is worth noting that the chemisorption energy of the OH groups for the configurations under consideration is approximately 1.2 to 1.6 eV/OH group higher than for nonmagnetic pairs[27,44] and lines.[51] This energy difference is smaller than the energies involved in the delamination process (on the order of 5 eV).

**MAGNETISM AND STABILITY OF CLUSTERS**

The computational results for the most stable configuration (Fig. 2c) are presented in Table 1. For the hydroxyl groups clusters, the total magnetic moment varied from 4.25 to 4.82 $\mu_B$ (depending on the presence of the ripple and the size of the supercell), which is in agreement with experimental data.[34] While the chemisorption of a single atom to graphene leads to the formation of quasilocalized (mid-gap) states,[10,49,52] the stronger distortion of graphene at the formation of the cluster from seven OH-groups leads to a smearing of this peak and local metallization (Fig. 4). To check for the possible influence of the electronic temperature and the k-point grid on the metallization of graphene, several calculations for different parameters were performed. The variation of the magnetic moment was not dependent (less than 0.05 $\mu_B$) on the choice of the electronic temperature (ranging from 10 to 300 K) or k-point number (ranging from 3 to 50). This explains the generation of a non-integer value for the magnetic moment; however, deviations from the "nominal" spin 5/2-state are not too large. For the clusters of hydrogen atoms at pristine graphene, the peak of the density of the states near the Fermi energy is much narrower (Fig. 4), and the magnetic moment of the cluster is therefore much closer to 5 $\mu_B$. At the optimized atomic structure for the hydrogen cluster on the ripple, reconstruction occurs (Fig. 2c) and the magnetic moment of the cluster drops to ½.

To estimate the exchange interaction between clusters, we calculated the energy difference between ferromagnetic (parallel magnetic moments) and antiferromagnetic (antiparallel magnetic moments) configurations of the clusters, and the energy difference decreased with the distance. At 2 nm, the energy is noticeable; however, at 3 nm, it has already become negligible (see Table 1). At the same time, the energy difference between the ferromagnetic and nonmagnetic configurations is weakly dependent on distance and is quite large, which is an intra-cluster characteristic.

To study the stability of the atomic configuration, we calculated the energy barriers for the migration of one OH-group in two directions, as shown in Fig. 2c. As we have previously shown,[27] the migration barriers are low if the distances between the OH-groups are less than 2 nm, which is due to long-range interactions between the hydroxyl groups on graphene[27,53-55] that are caused by long-range hydrogen bonds[56] and significant distortions of the graphene sheet within a of radius of 2 nm.[17] When distance decreases, the barriers become lower and reconstruction to a more energetically favorable nonmagnetic configuration occurs. Therefore, only the clusters with a distance between them of more than 3 nm can be a potential source of large magnetic moments. Their magnetic state is only metastable; however, it may be protected by high-energy barriers.[57,58]

For chemisorption at a graphene bilayer, atomic distortions of the upper layer are less than the distortions for a single layer of graphene.[25] It leads to the disappearance of the barrier for the motion of the OH group to the center of cluster (see Table 1). Thus, one can conclude that high-spin clusters of the type considered here are unstable in multilayer graphene and graphite. In the case of single-layer graphene, when the distance between the clusters increases, the barrier for migration inside the cluster grows and a barrier for migration outside the cluster is generated.

The barrier for OH clusters in flat graphene is approximately 0.4 eV (see Table 1), which is not enough to protect their stability for a long time at room temperature (they would be long-lived if they formed at temperatures below 100 K). However, the barrier will be higher if the cluster is formed on top of a ripple. A ripple with a height of approximately 1 Å (Fig. 1c) suppresses this interaction by changing the orientation (Fig. 1b). For the chosen parameters, which were more or less arbitrary, the energy barrier increased up to 0.7 eV in our simulations. This is close to the energy barrier for migration of magnetic mono-vacant graphite,[59] which is considered as a potential source of room-temperature magnetism in irradiated graphite.[60] There is not enough information on the experimental parameters of the ripples[61-64] in graphene laminate; therefore, one can expect even higher barriers and a high-spin configuration that will be long-lived at room temperature.

For the hydrogen clusters, there were no hydrogen bonds and lattice distortions that were smaller than the OH groups. The lack of hydrogen bonds and lattice distortions made the energetics of the clusters different. The high-spin state of hydrogen appears to be unstable (see Table 1). Experimentally, paramagnetism with high spins was observed when DMF was used as the solvent and was never observed when NMP was used.[45] Our model is in agreement with the observation for NMP because only hydrogen clusters can be formed. In addition, we predict that the use of chloroform as a solvent will not be paramagnetic; however, this has not been validated.

## CONCLUSIONS

Our calculations show that migration of the hydroxyl groups from the solvent to form magnetic configurations during the production graphene laminate is possible if the ripples on graphene are involved. The most stable magnetic configuration corresponds to seven OH-groups and reflects the symmetry of the graphene lattice (Figs. 1 and 2c). The total magnetic moment of these clusters is between 4 and 5 $\mu_B$, which is in agreement with the experimental results.[34] This configuration is special because clusters of smaller or larger sizes, as well as less-symmetric clusters, are much less energetically favorable and stable. The structural stabilization of the clusters is possible if the distance between the clusters is larger than 3 nm. Under this condition, the exchange interaction between different clusters is negligible and they can be considered free paramagnetic centers, which is in agreement with the experimental findings.[34]

## METHODS

The modeling was performed by density functional theory (DFT) in the pseudopotential code SIESTA,[65] as was done in our previous works.[18,26-28,44,50] All calculations were performed using the generalized gradient approximation (GGA-PBE),[66] which is the most suitable approximation for the description of graphene-adatom chemical bonds.[27] Full optimization of the atomic positions was performed. During the optimization, the electronic ground state was consistently found using norm-conserving pseudo-potentials for cores, a double-ζ plus polarization basis of localized orbitals for carbon and oxygen, and a double-ζ basis for hydrogen. Optimization of the force and total energy was performed with an accuracy of 0.04 eV/Å and 1 meV, respectively. All calculations were carried out with an energy mesh cut-off of 360 Ry and a k-point mesh of 6×6×1 in the Mokhorst-Park scheme.[67] When drawing the pictures of the density of states, a smearing of 0.2 eV was used. All calculations were performed in the spin-polarized mode. Values of the energy barriers were calculated according to the method described in the work: (i) first, the full optimization of atomic structures was performed for the initial and final points of migration; (ii) several points homogeneously distributed were chosen along the migration path; and (iii) the whole optimization of the atomic structure for these points was performed with fixed coordinates for the migrated atom or atomic group. The migration barrier was defined as the total energy difference between the initial point and the intermediate point with the highest energy along the migration path.[27] Migration without barrier indicates that the initial point corresponds to the total energy maximum.

The chemisorption energy and stability of the chemisorbed clusters were sensitive to the local curvature of the graphene layer.[50] Therefore, we have performed calculations for both flat graphene (Fig. 1a) and rippled graphene (Figs. 1b and c). Ripples with a height of 1.6 Å and a radius of 7.9 Å were chosen for the initial iteration. Optimization yielded a ripple structure with a height of approximately 1 Å and a radius of 4.5 Å for all chemisorbed species studied.

Because the migration barriers strongly depended on the distance between the adsorbed hydroxyl groups, we have chosen two supercells of graphene, one that was 8×8×1 and another that was 12×12×1, which contained 128 and 288 carbon atoms, respectively. In the first case, the distance between the centers of the cluster is approximately 2 nm, with 1 nm between neighboring hydroxyl grounds within the cluster, and, in the second case, 3 and 2 nm, respectively. The graphene laminate used in the experiment[34] contained a noticeable fraction of bilayer graphene; therefore, we have considered the case of a bilayer with a supercell containing 128 carbon atoms per layer. To calculate exchange interaction parameters, we increased the size of the supercell with the optimized atomic structure along one of the directions of the XY-plane (8×16×1 and 12×24×1 instead of 8×8×1 and 12×12×1, respectively) and calculated the total energy for the parallel and antiparallel orientations of the spins from two magnetic clusters.

*Acknowledgements.* We are grateful to Andre Geim and Irina Grigorieva for stimulating discussions. MIK acknowledges support from the "Stichting voor Fundamenteel Onderzoek der Materie (FOM)," which is financially supported by the "Nederlandse Organisatie voor Wetenschappelijk Onderzoek (NWO)."

**Table I.** Values of the magnetic moment per system, the energy differences between magnetic and non-magnetic configurations, ferromagnetic and antiferromagnetic configurations, and the energy barrier for the migration of a single hydroxyl group or hydrogen adatom (in parenthesis) to the center of cluster (in) or out of the center of the cluster (see Fig. 1b).

| formulas | $C_{128}(OH)_7$ ($C_{128}H_7$) | | $C_{256}(OH)_7$ ($C_{256}H_7$) | $C_{288}(OH)_7$ ($C_{288}H_7$) | |
|---|---|---|---|---|---|
| number of layers | 1 | | 2 | 1 | |
| corrugation | flat | ripple | flat | flat | ripple |
| magnetic moment, $\mu_B$ | 4.62 (4.99) | 4.82 (unstable) | 4.25 (4.96) | 4.30 (4.96) | 4.45 (unstable) |
| $E_{FM}-E_{nonmag}$, meV | 248 (641) | 306 (unstable) | 184 (547) | 221 (488) | 227 (unstable) |
| $(E_{FM}-E_{AFM})/2$, meV | 86 (194) | 126 (unstable) | 88 (170) | 4 (17) | 2 (unstable) |
| Energy barrier in/out, meV | 362/--- (96/---) | 387/--- (unstable) | ---/--- (---/---) | 496/372 (110/---) | 680/1023 (unstable) |

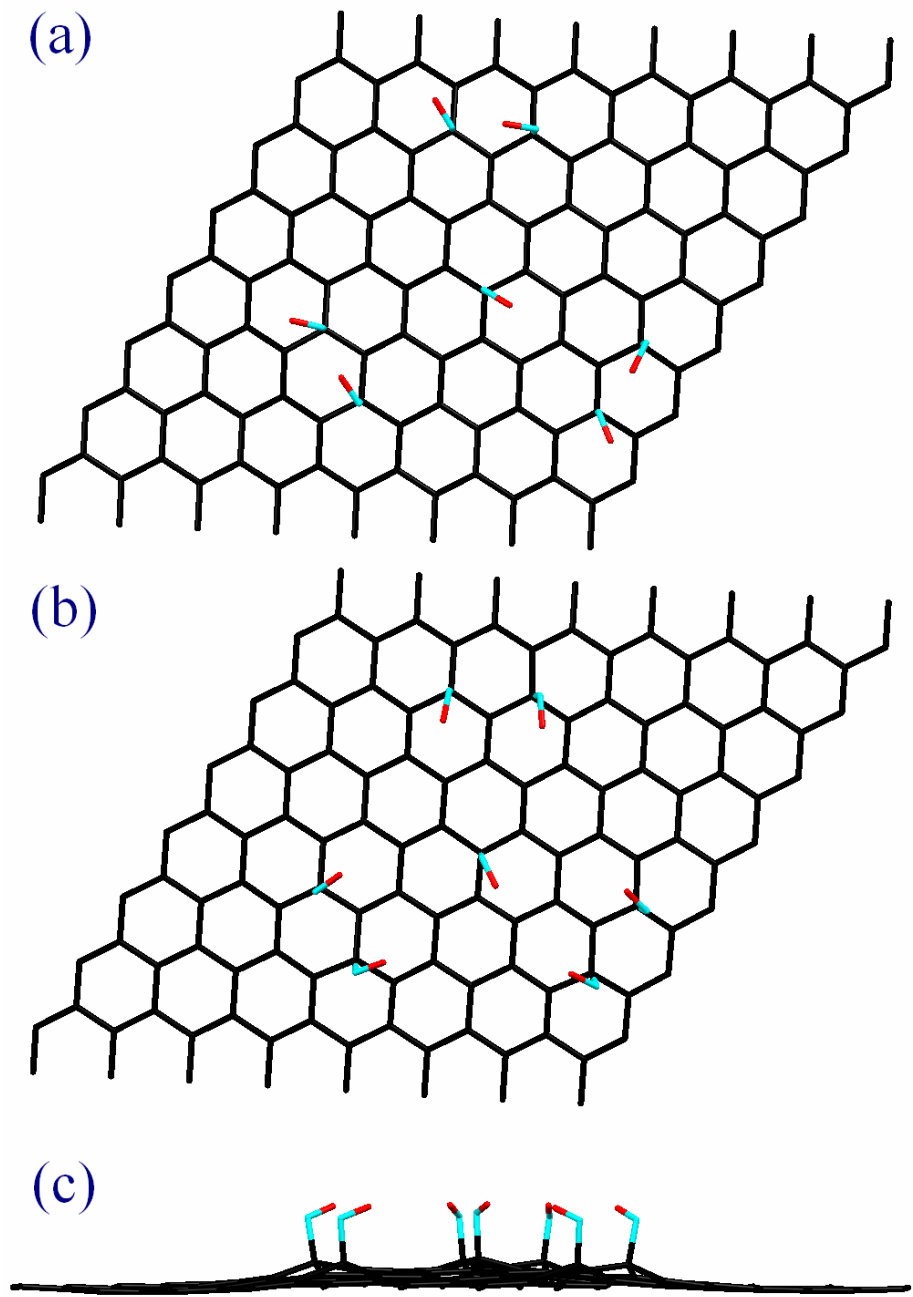

**Fig. 1** Optimized atomic structures of the system $C_{128}(OH)_7$ for flat graphene (a) and a ripple (b and c). Carbon atoms are shown in black, hydrogen atoms are red and oxygen atoms are shown in cyan.

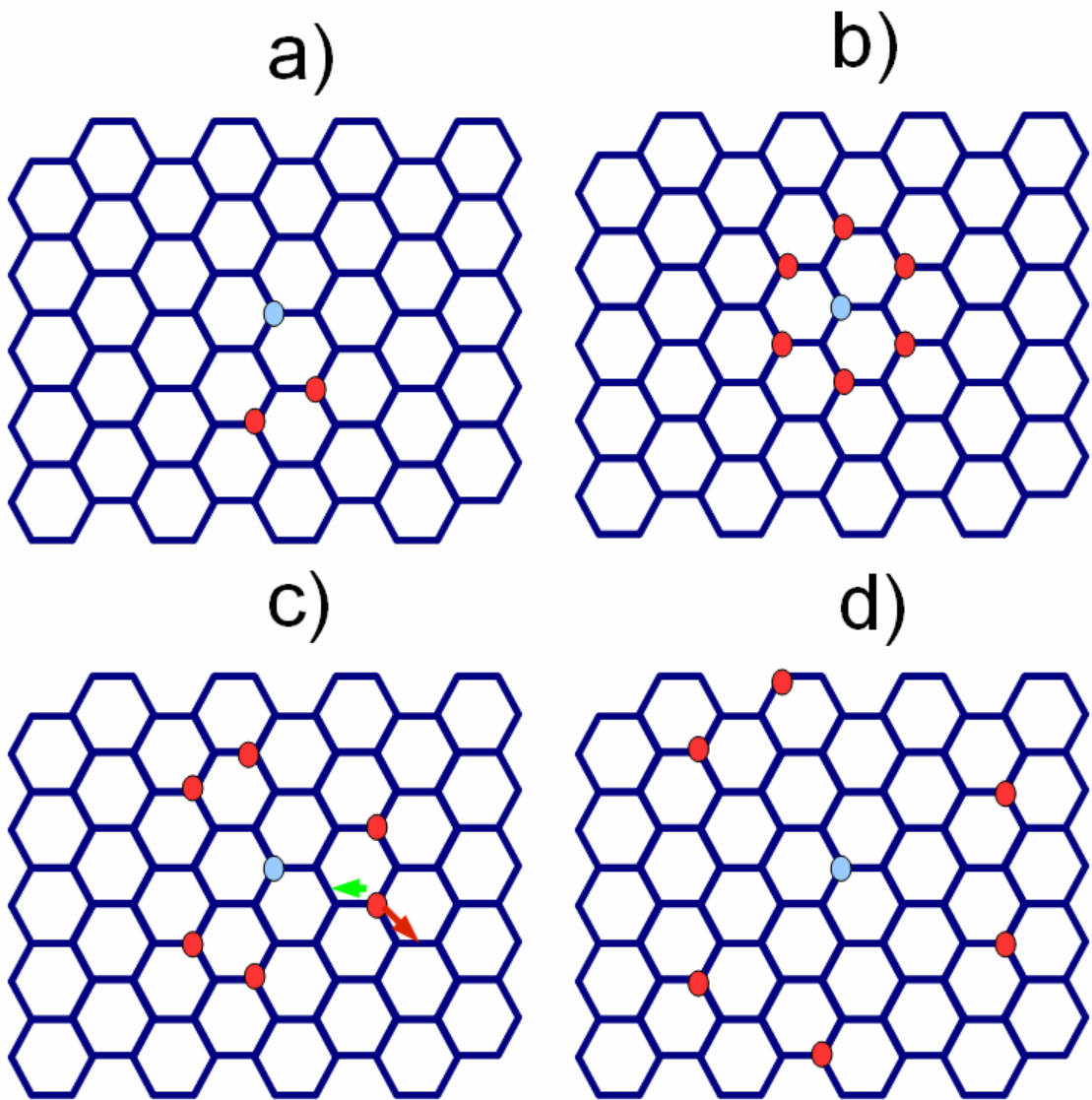

**Fig. 2** Structures of magnetic clusters with a total magnetic moment (in localized-spin approximation) of 1 $\mu_B$ (a) and 5 $\mu_B$ (b-d), respectively. The light blue and red colors represent the hydroxyl groups chemisorbed in different sublattices. Green and red arrows in the panel (c) show possible migrations for one of the hydroxyl groups (see the text).

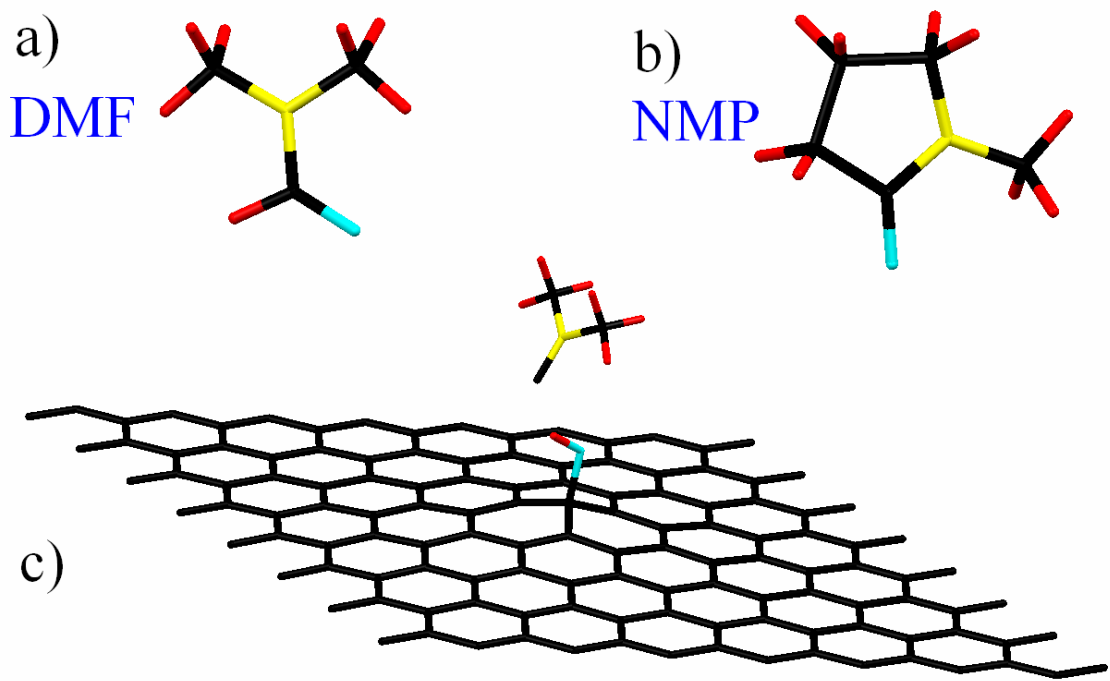

**Fig. 3** Optimized atomic structure of (a) dimethylformamide (DMF) and (b) methylpyrrolidone (NMP) molecules used as solvents for graphene production, and migration of hydrogen and oxygen atoms from a DMF molecule to a graphene sheet with hydroxyl group formation (c). Carbon atoms are shown in black, hydrogen atoms are red, oxygen atoms are cyan, and nitrogen atoms are shown in yellow.

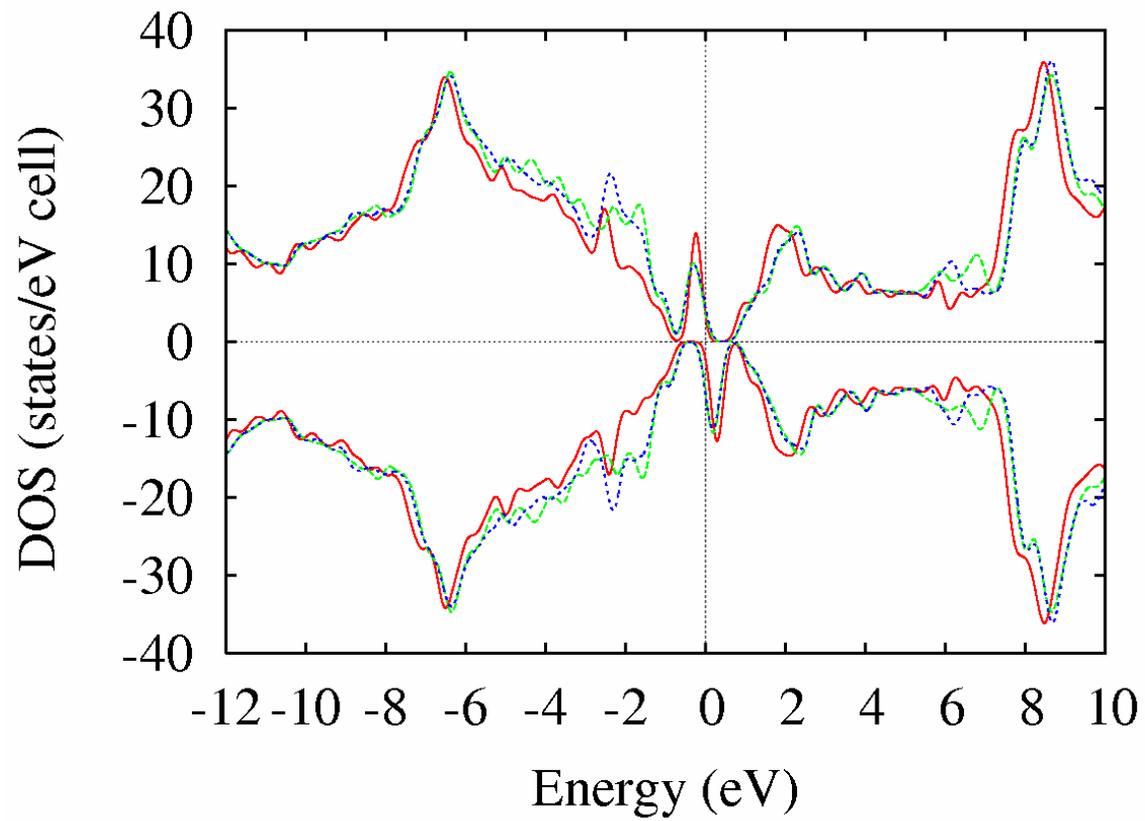

**Fig. 4** Total density of states for the systems $C_{128}H_7$ (solid red line) and $C_{128}(OH)_7$ for flat graphene (dotted blue line) and graphene with a ripple (dashed green line).